# Self-Supervised Learning from Unlabeled Fundus Photographs Improves Segmentation of the Retina

**Jan Kukačka**[* 1,2], **Anja Zenz**[* 1,2], **Marcel Kollovieh**[1,2,3], **Dominik Jüstel**[1,2,3], and **Vasilis Ntziachristos**[1,2,✉]

## Abstract

Fundus photography is the primary method for retinal imaging and essential for diabetic retinopathy prevention. Automated segmentation of fundus photographs would improve the quality, capacity, and cost-effectiveness of eye care screening programs. However, current segmentation methods are not robust towards the diversity in imaging conditions and pathologies typical for real-world clinical applications. To overcome these limitations, we utilized contrastive self-supervised learning to exploit the large variety of unlabeled fundus images in the publicly available EyePACS dataset. We pre-trained an encoder of a U-Net, which we later fine-tuned on several retinal vessel and lesion segmentation datasets. We demonstrate for the first time that by using contrastive self-supervised learning, the pre-trained network can recognize blood vessels, optic disc, fovea, and various lesions without being provided any labels. Furthermore, when fine-tuned on a downstream blood vessel segmentation task, such pre-trained networks achieve state-of-the-art performance on images from different datasets. Additionally, the pre-training also leads to shorter training times and an improved few-shot performance on both blood vessel and lesion segmentation tasks. Altogether, our results showcase the benefits of contrastive self-supervised pre-training which can play a crucial role in real-world clinical applications requiring robust models able to adapt to new devices with only a few annotated samples.

## 1   Introduction

Fundus photography, the primary method for retinal imaging [1], allows early detection of both eye and systemic diseases [1-4]. Thanks to the advances in telemedicine and emergence of cheap fundus cameras [5-7], fundus photography plays the key role in scaling up diabetic retinopathy prevention in developing countries [8-12]. However, due to lack of qualified ophthalmologists to diagnose the images [12-14], automated methods are needed to manage the growing volume of fundus photography examinations [15-17]. In particular, manual segmentation of retinal blood vessels is prohibitively time-consuming, and its automation would facilitate identifying and quantifying biomarkers such as vessel caliber [18, 19], tortuosity [20], and branching geometry [21, 22], which would support assessment of conditions such as diabetic retinopathy [23] and cardiovascular disease [3, 4]. Furthermore, automatic segmentation of retinal lesions [24] and other anatomical targets [25] could yield additional biomarkers.

[1] Helmholtz Zentrum München (GmbH), Neuherberg, Germany, Institute of Biological and Medical Imaging. [2] Technical University of Munich, Germany, School of Medicine, Chair of Biological Imaging, Center for translational Cancer Research (TranslaTUM). [3] Helmholtz Zentrum München (GmbH), Neuherberg, Germany, Institute of Computational Biology. [*] Authors contributed equally to this work. [✉] email: v.ntziachristos@tum.de

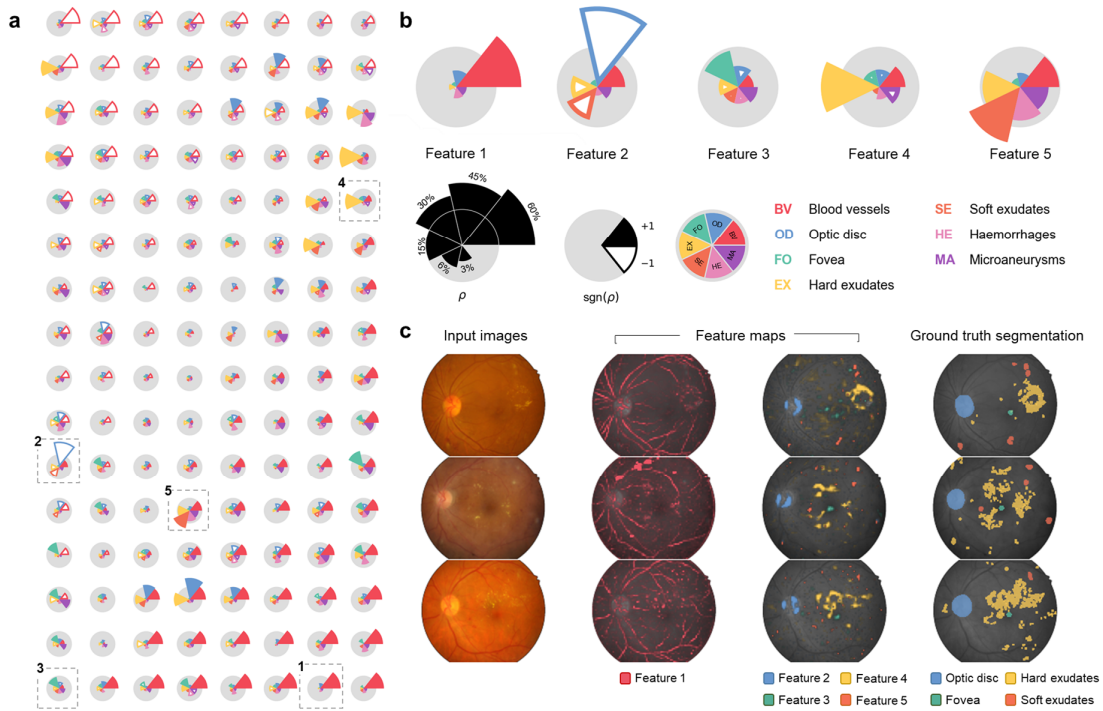

**Fig. 1**: Visualization of the features learned by the self-supervised training. **a)** Diagram of Pearson correlation coefficients between the 128 learned features and localization of seven anatomical and pathological targets: blood vessels, optic disc, fovea, hard exudates, soft exudates, haemorrhages, and microaneurysms. **b)** Enlarged diagrams for five features marked in A. **c)** Three representative images from the IDRiD dataset (left), corresponding activations maps of features 1–5 (middle two columns), and ground truth segmentations of optic disc, fovea, hard exudates, and soft exudates (right). Visual comparison of columns 3 and 4 shows a good spatial match between activations of these selected units with true segmentation.

State-of-the-art approaches for automated retinal segmentation rely on supervised deep learning [26]. Achieving great performance on standardized, consistent datasets with even small amounts of training data (10–20 images), the performance of these methods drops considerably when exposed to changes in imaging conditions—i.e., *domain shift* [27]. To mitigate this issue, the model robustness can be slightly improved by fine-tuning it on new data, using its own predictions as pseudo-labels [27]. The requirement for re-training of the model on the target data reduces the practicality of this method, however. Alternatively, labeled data from multiple datasets can be combined to increase the robustness to domain shift [28], but the scalability of this approach is limited by the variety of labeled data available. Ultimately, as the variety of fundus images keeps growing, it is becoming increasingly difficult to create a labeled dataset sufficiently representing all different cameras, diseases, and imaging conditions.

A potential new direction for overcoming the need for labeled data is self-supervised learning, where neural networks are trained on an auxiliary *pre-text task*. Pre-text tasks are designed such that targets can be easily generated from unlabeled data and solving them requires learning a semantic representation of the data. This representation can be further fine-tuned in the usual supervised manner on a *downstream task*. Self-supervised pre-training employing an image restoration pre-text task has been demonstrated for radiological images, outperforming generic representations transferred from ImageNet classification [29]. In the domain of fundus images, several self-supervised pre-training strategies have been shown to improve the accuracy of diabetic retinopathy classification, such as generic visual tasks performed on unlabeled images [30] or depth prediction from matching optical coherence tomography (OCT) images [31]. A multi-modal approach using fluorescein angiography images has led to enhanced optical cup and disc segmentation [32].

Recently, using the pre-text task of contrastive instance discrimination has become the primary method for self-supervised learning of general visual representations from large datasets such as the ImageNet [33-35]. In this type of learning, a neural network is trained to map randomly transformed variants of input images to a new representation, where samples originating from the same image are similar to each other and dissimilar to samples from other images. Despite its success in the general vision domain, the applicability of contrastive self-supervised learning to retinal imaging has not been demonstrated yet. We hypothesized that when combined with a rich, large dataset of retinal photographs such as the EyePACS dataset [36], the method can be used to learn a useful domain-specific representation from unlabeled images. Such a representation could improve the performance on downstream segmentation tasks.

Herein, we demonstrate for the first time that a convolutional neural network trained in the contrastive self-supervised manner on the EyePACS dataset learned to recognize retinal anatomical structures and various types of lesions without being provided any labels. Moreover, the learned image representation is robust to domain shift owing to the large variety of the dataset and data augmentation used during the training. Using this representation as an encoder of a U-Net, we observed three main benefits for downstream retinal segmentation over a baseline U-Net trained from random initialization: 1) the pre-trained model is more robust to domain shift and outperforms the baseline when applied to data from a different dataset (with different population and imaging conditions) in 4 out of 6 experiments, achieving state-of-the-art performance in three cases; 2) the pre-trained model achieves better overall performance than the baseline in scenarios with limited labeled images available; 3) the pre-trained model requires significantly less epochs than the baseline to reach its peak validation set performance on blood vessels, hard exudates, and hemorrhage segmentation tasks.

Overall, our work presents the capability of contrastive self-supervised learning to exploit available unlabeled fundus images and turn them into a domain-specific representation beneficial for automated segmentation of retinal structures.

## 2 Results

We performed three experiments: 1) the representation learning experiment to characterize the representation obtained through contrastive self-supervised learning, 2) the image segmentation experiment to evaluate the benefits of using the pre-trained representation for retinal lesion and vessel segmentation, and 3) the domain transfer experiment to evaluate if the pre-trained models are robust and transfer well across different vessel segmentation datasets. Details of the experiments are listed in Methods.

### 2.1 Representation learning experiment

Our first experiment revealed that despite using no labels, the representation obtained by self-supervised learning contains features that are specialized in recognizing various retinal structures. Specifically, we evaluated the spatial correlation of seven anatomical and pathological structures with the activation maps in the last layer of an encoder trained on unlabeled images from the EyePACS dataset. Fig. 1a shows that many of the 128 units developed a strong correlation with blood vessels, but some specialized in detecting other structures. Fig. 1b shows a detailed correlation of five selected units that respond strongly to blood vessels, optic disc, fovea, hard exudates, and soft exudates. Fig. 1c displays the spatial agreement of the activations of these five units with the respective targets. Good preservation of the spatial information is an essential feature of the self-

**Table 1:** Performance (AUPRC, %) of the best methods for retinal lesion segmentation from the IDRiD Grand challenge leaderboard (sub-challenge 1). Top two rows show test set performance of our U-Net trained with random initialization (baseline) and with self-supervised pre-training. Bold highlights highest performance in each column. Our experiments are reported as mean ± standard deviation of four runs with different training/validation splits. For performance of other methods, we report scores from the challenge leaderboard evaluated on the same test set as our results [37].

| Method | EX | SE | HE | MA |
|---|---|---|---|---|
| Random initialization | 88.33 ± 1.46 | 67.43 ± 1.82 | 64.73 ± 1.12 | 50.70 ± 1.37 |
| Self-supervised pre-training | 89.42 ± 0.60 | 69.98 ± 5.46 | 66.37 ± 1.72 | 48.42 ± 1.87 |
| PATech | 88.50 | - | 64.90 | 47.40 |
| VRT | 71.27 | 69.95 | 68.04 | 49.51 |
| iFLYTEK | 87.41 | 65.88 | 55.88 | 50.17 |
| LzyUNCC-I | 76.15 | 66.07 | - | - |
| LzyUNCC-II | 82.02 | 62.59 | - | - |
| SAIHST | 85.82 | - | - | - |
| SOONER | 73.90 | 53.69 | 53.95 | 40.03 |

EX = hard exudates, SE = soft exudates, HE = hemorrhages, MA = microaneurysms.

supervised representation for precise downstream segmentation. Additionally, the features are detected in images of a different population from a separate dataset (US vs India), demonstrating the robustness of the representation to domain shifts.

## 2.2 Image segmentation experiment

Our second experiment confirmed that using the self-supervised representation as an initialization of the encoder of a U-Net benefits the downstream segmentation tasks and leads to higher segmentation accuracy, especially in the few-shot regime, and faster convergence.

First, we compared the performance of a pre-trained U-Net to a randomly initialized baseline on the IDRiD retinal lesion segmentation dataset and observed an improvement in three out of four lesion types. Table 1 shows that the pre-trained model achieved on average a higher AUPRC for hard exudates (EX; 1.09 percentage points, one-sided 95% CI [0.05, $+\infty$]), soft exudates (SE; 2.55 pp), and hemorrhages (HE; 1.64 pp, [0.29, $+\infty$]), but not for microaneurysms (MA; $-2.27$ pp [$-5.24$, $+\infty$]). Furthermore, the table lists reported results of the best performing methods from the IDRiD Grand challenge [37]. For EX and SE, the pre-trained model outperformed the challenge winners, whereas for HE and MA its mean performance was within 1 standard deviation of the best method. To provide a qualitative illustration of predictions by the pre-trained model, Fig. 2a shows a representative example of the segmentation.

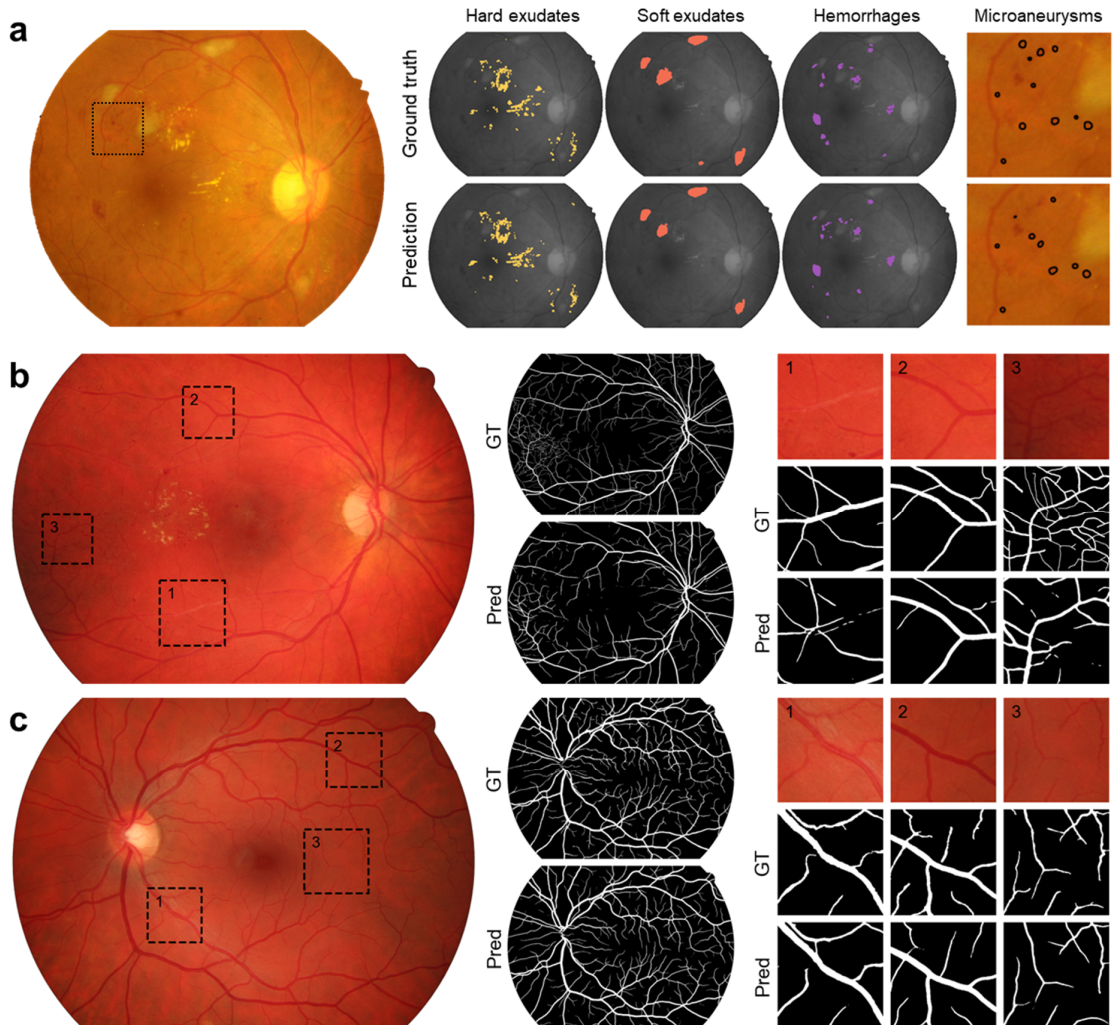

**Fig. 2:** Examples of lesion and blood vessel segmentation quality achieved by the pre-trained models. **a)** Image #59 from IDRiD test set, ground truth segmentations, and segmentation predictions. Microaneurysms are displayed on an enlarged cut-out due to their small size (enlarged region marked in the left image). **b)** Diabetic retinopathy patient #6 of HRF dataset (worst case, Dice 72.84%). **c)** Healthy patient #12 of HRF dataset (best case, Dice 88.30%). Cut-outs showing arteries (1), veins (2), and microvessels (3) with corresponding ground truth and predicted segmentation. Locations of the cut-outs are delineated in the images on the left in panels b and c.

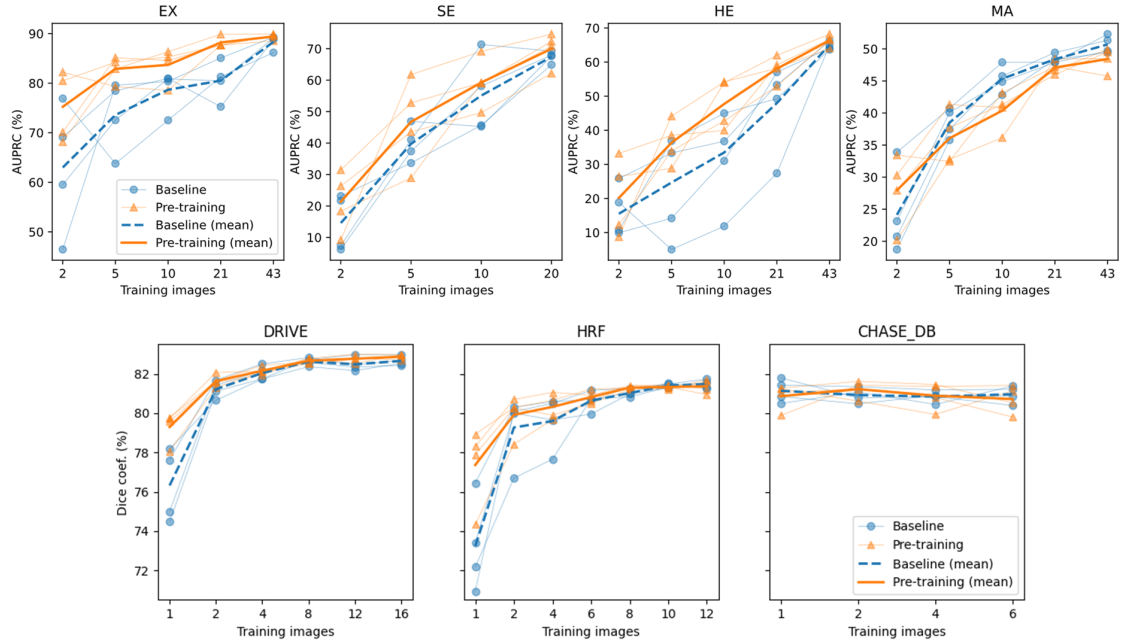

**Fig. 3**: Performance comparison of pre-trained and randomly initialized U-Net on IDRiD lesion segmentation dataset (upper row) and retinal vasculature segmentation datasets DRIVE, HRF, and CHASE-DB (bottom row) for decreasing amount of training data. The experiments were repeated four times with different training/validation splits (individual runs are plotted as thin lines, thick lines represent the means). EX = hard exudates, SE = soft exudates, HE = hemorrhages, MA = microaneurysms.

Fig. 3 shows that the pre-trained model outperformed the baseline even more when we reduced the number of training images. The improvements were statistically significant for hard exudates (21 images: 7.69 pp [2.74, +∞], 10 images: 5.02 pp [3.97, +∞], 5 images: 9.31 pp [3.77, +∞], 2 images: 12.25 pp [2.83, +∞]) and hemorrhages (10 images: 16.56 pp [0.91, +∞]). However, for microaneurysms, the pre-trained model performed better only in the case of 2 training images.

Second, we compared the performance of a pre-trained U-Net to a randomly initialized baseline on the three retinal vessel segmentation datasets, but in this case, we did not observe any significant differences. Table 2 shows that the Dice score on DRIVE improved slightly (0.21 pp, two-sided 95% CI [−0.21, 0.63]) and decreased a little on CHASE-DB (−0.25 pp [−1.13, 0.64]) and HRF (−0.11 pp, [−0.40, 0.19]). As the confidence intervals show, neither of the differences is statistically significant. Additionally, the table lists the performance reported by several state-of-the-art methods (asterisks mark deviations from our evaluation protocol; see Methods, 4.6). Our results confirm that a simple U-Net achieves comparable results to more complicated, specialized architectures.

**Table 2:** Performance (Dice coefficient, %) of retinal vasculature segmentation on DRIVE, CHASE-DB, and HRF. Our results are reported as mean ± std. deviation of four runs with different training/validation splits. For other methods we list results reported by their authors. We followed exactly the same evaluation procedure, including testing split, described in [27]. Methods using different training/testing split or different score computation procedure are marked by an asterisk (*).

| Method | DRIVE | CHASE-DB | HRF |
|---|---|---|---|
| Random initialization | 82.66 ± 0.20 | 80.97 ± 0.44 | 81.48 ± 0.22 |
| Self-supervised pre-training | 82.87 ± 0.13 | 80.72 ± 0.71 | 81.38 ± 0.35 |
| Little W-Net [27] | 82.82 | 81.55 | 81.04 |
| M-GAN [38, 39] | 83.17 ± 0.02 | 81.10 | 79.72* |
| HAnet [40] | 82.93 | 81.91* | 80.74 |
| M2U-Net [28] | 80.30 ± 1.42* | 80.22 ± 1.93* | 78.00 ± 5.74* |
| VGN [41] | 82.63* | 80.34* | 81.51* |
| DEU-Net [42] | 82.70 | 80.37* | |
| DUNet [39] | 82.37 | 78.83* | |

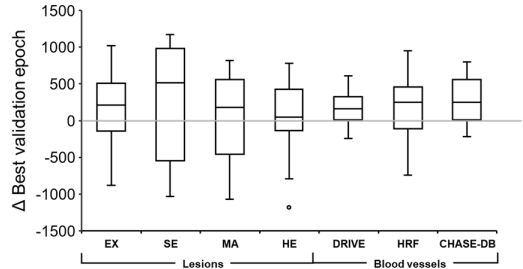

**Fig. 4**: Training length improvement of pre-trained network over a randomly initialized U-Net.

To provide representative segmentation examples, Fig. 2b and c show the worst and the best images from the HRF dataset respectively, with detailed segmentations of arteries (1), veins (2), and microvessels (3). Fig. 2b shows a diabetic retinopathy case where the model achieved a 72.84% Dice score. Here, the model correctly segmented the veins but failed to detect arteries with strong central reflex lending them a whitish appearance and atypical retinopathic neovascularization in a dark region of the image. Fig. 2c shows a healthy case where the model provided a correct segmentation for veins, microvessels, and arteries even in the presence of central reflex, and achieved a Dice of 88.30%.

In the few-shot regime, the pre-trained model outperformed the baseline on the DRIVE and HRF datasets but not on CHASE-DB. Fig. 3 shows that the gap between the models grows as we reduce the number of available training images. For DRIVE, we observed significant improvement over baseline in case of one training image (2.98 pp, one-sided 95% CI [0.11, +∞]) and two training images (0.42 pp [0.01, +∞]). For HRF, the improvement was significant in case of one training image (4.12 pp [0.90, +∞]) and eight training images (0.28 pp [−0.03, +∞]). The segmentation task for the CHASE-DB dataset appears easier than in DRIVE and HRF and reducing the number of training images down to a single one does not alter the performance of either model.

Besides improved segmentation performance, the pre-trained models were also converging faster. Fig. 4 shows the distribution of differences in the number of training epochs needed by the baseline and the pre-trained model to reach the best validation error across the lesion and blood vessel segmentation tasks. Supplementary table 4 summarizes the exact differences and shows that, except for soft exudates and microaneurysms, the improvements are statistically significant.

### 2.3 Domain transfer experiment (cross-training)

Our final experiment demonstrated that pre-trained models are also more robust to domain shift and thus better suited for domain transfer. Table 3 shows a performance comparison of models trained on DRIVE dataset and tested on other retinal vasculature segmentation datasets. In four of the six cases (STARE, LES-AV, DR HAGIS, AV-WIDE), the pre-trained model outperformed the baseline. The improvement was statistically significant for DR HAGIS (0.62 pp, [0, +∞]) and AV-WIDE (1.89 pp, [0.14, +∞]). Furthermore, the table shows that in three cases (HRF, DR HAGIS, and AV-WIDE), our approach also outperformed Little W-Net [27], a state-of-the-art approach in retinal segmentation domain transfer using pseudo-label re-training on the target dataset.

## 3 Discussion

In this work, we have demonstrated for the first time the feasibility of learning robust representations of fundus images capable of recognizing important anatomical and pathological features in a completely label-free manner. Furthermore, we have identified benefits of using this representation for downstream image segmentation tasks, notably, an improved performance in few-label scenarios, shorter training times, and improved robustness to domain shifts. Our results suggest contrastive self-supervised learning as an effective way to exploit unlabeled fundus images and advance automated retinal diagnostics.

**Table 3**: Cross-training evaluation. Dice coefficients (%). Mean ± standard deviation of 12 runs is reported. For Little W-Net, results reported by its authors are shown.

| Method | CHASE-DB | HRF | STARE | LES-AV | DR HAGIS | AV-WIDE |
|---|---|---|---|---|---|---|
| Random init. | 75.91 ± 0.61 | 75.93 ± 0.41 | 78.47 ± 0.62 | 77.12 ± 1.12 | 70.91 ± 1.16 | 67.43 ± 3.70 |
| SSL pre-training | 75.79 ± 0.52 | 75.84 ± 0.51 | 78.77 ± 0.65 | 77.31 ± 0.79 | 71.54 ± 1.09 | 69.32 ± 1.51 |
| Little W-Net [27] | 76.49 | 71.12 | 79.76 | 77.93 | 68.67 | 62.46 |

Various methods to obtain useful representations of fundus images have been considered in the past. Unsupervised methods relying on handcrafted features have long been used to segment retinal vasculature [43-45]. It was recently shown that weakly supervised convolutional neural networks implicitly learn to detect retinopathic lesions [46, 47]. Our work is the first demonstration that a conv-net can learn to recognize retinal vasculature and lesions in a fully data-driven manner without any labels. By learning directly from the data, the resulting representation is more robust than handcrafted approaches. Moreover, requiring no labels, the self-supervised approach can be easily applied to unannotated datasets, which are cheaper to obtain.

Our work provides insight into the features learned during self-supervised pre-training. Currently, self-supervised pre-training is mostly used in other domains as a black-box method, and there is very little insight into the properties and quality of the learned representation. It is well recognized that the representation depends on numerous hyperparameters, such as stochastic transformations used for generating training samples and the encoder architecture, but their immediate effect on downstream tasks is not clear [33, 48, 49]. Only few studies have inspected the qualitative properties of the self-supervised representations, such as exploring nearest neighbors in the embedding space [49, 50], visualizing saliency maps [49], reconstructing inputs from the embeddings [51], or studying the intrinsic dimensionality of the embedded dataset [50]. Similar to these approaches, our feature correlation analysis (Fig. 1) allows to easily assess the impact of changing the hyperparameters on the features that the network recognizes without lengthy fine-tuning on a downstream task.

Furthermore, we have identified specific benefits of using the self-supervised representation for downstream segmentation tasks. First, it achieved better segmentation performance in a few-shot learning regime than a baseline trained from scratch. In vessel segmentation tasks, this improvement diminished when we sufficiently increased the number of labeled images (Fig. 3). For more difficult, lesion segmentation tasks (EX, SE, HE), we did not observe such saturation even when the full training dataset was used (Fig. 3). The same saturation pattern was also recently reported on simulated data [52]. An interesting exception was performance on the CHASE-DB dataset, where the performance with full training dataset was comparable to performance of using a single training image for both baseline and pre-trained models. This task requires segmenting only large vessels and appears to be straightforward. Another exception was segmentation of microaneurysms where the pre-trained model performed consistently worse than the baseline. We assume this is because the pre-trained model does not contain any features that correlate very well with MA (Fig. 1) and cannot adapt to new targets from its initial configuration. Using different data augmentation enhancing their contrast could improve this situation.

The second observed benefit was faster convergence of pre-trained models. This is not surprising since the pre-trained model has already learned necessary low-level feature detectors and similar results were reported in other studies [30, 53]. In practice, faster convergence can lead to large computation savings for hyperparameter tuning on the downstream task and serves as amortization of the pre-training computation costs.

The third benefit was improved robustness to domain shifts—probably the most unique and important advantage of using self-supervised pre-training (Table 3). Self-supervised pre-training outperforms the pseudo-label method of Little W-Net in three out of six cases [27] but is generally worse than training on a combination of multiple datasets [28]. As these three approaches are independent, we expect that using them in combination could lead to superior results. We also remark that in cases of HRF, DR HAGIS, and AV-WIDE, we resized the target images to a width of 1024 px instead of 512 px used by the Little W-Net authors, which granted quite large improvements (observable in our baseline performance, Table 3). We conclude that attention must be paid to proper resolution matching between source and target images, taking into account the field-of-view of the cameras and aspect-ratio of the image files, since a resolution mismatch can easily hinder improvements granted by advanced machine learning techniques.

The advantages of self-supervised pretraining are essential for utilization in fast-evolving, low-cost, smartphone-based fundus cameras, which exhibit large variability in imaging conditions and require methods that can easily adapt to new devices without large, annotated datasets. Additionally, emerging medical imaging modalities, lacking large standardized labeled datasets, can also benefit from this type of representation learning. Moreover, unlike large-scale self-supervised visual models, our approach can be easily used with customer-level GPUs (pre-training for 600 epochs took 7 hrs on a single NVidia RTX 3090).

Whereas our experiments show promising results, open questions remain. It has been shown that large self-supervised networks can learn strong representations from huge datasets with billions of images and outperform their fully supervised counterparts [35]. Conversely, the representation learned by our small network did not improve beyond 21k images (only 1/4 of the available data). Using a larger encoder might enable learning even better representations. On the other hand, huge networks often perform poorly on small downstream datasets typical for the medical domain, and careful layer freezing [53] or knowledge distillation [35, 54] might be necessary to prevent overfitting.

Overall, this work demonstrates how contrastive self-supervised learning can be applied to segmentation of fundus photographs. We identified specific benefits, such as better few-shot performance, faster convergence times, and improved domain transfer. These benefits are relevant for deploying automated models into screening programs based on fundus photography, which is essential to reduce the workload on individual ophthalmologists and increase the capacity of current healthcare systems.

**Funding.** This project has received funding from the European Research Council (ERC) under the European Union's Horizon 2020 research and innovation programme under grant agreement No 694968 (PREMSOT).

**Acknowledgments.** The authors would like to thank Sergey Sulima for assistance in preparing this manuscript.

**Disclosures.** Vasilis Ntziachristis is an equity owner and consultant at iThera Medical GmbH, member of the Scientific Advisory Board at SurgVision BV / Bracco Sp.A, owner at Spear UG, founder and consultant at I3, and founder of Sthesis.

**Data availability.** Code, trained models, and full results are publicly available at `github.com/jankukacka/ssl_retinal_segmentation`

# 4 Methods

## 4.1 Experiments

To evaluate the potential and benefits of contrastive self-supervised learning for retinal segmentation, we performed three experiments.

**Representation learning experiment.** We trained a convolutional encoder in a self-supervised manner on unlabeled images from the EyePACS dataset. Then, we inspected the learned representation by correlating the output feature activation maps to ground truth segmentation of retinal anatomical and pathological targets in the DRIVE, IDRiD, and HRF datasets to identify which targets are recognized by the network.

**Image segmentation experiment.** We appended a decoder to the trained encoder to form a U-Net network for image segmentation. We fine-tuned this network on DRIVE, HRF, and CHASE-DB retinal vasculature segmentation datasets and IDRiD retinal lesion segmentation dataset and compared its performance to a baseline U-Net trained from scratch. To examine the benefits of a pre-trained encoder over the baseline in limited data scenarios, we repeated this experiment with decreasing numbers of training images down to a single one. We compared the two approaches in terms of segmentation quality and the number of training epochs needed to reach the optimal performance. To increase the sample size for the latter comparison, we pooled together experiments with varying numbers of training images.

**Domain transfer experiment** (cross-training). To evaluate if pre-training improves the robustness of the model to domain shift, we repeated the cross-training experiment of Galdran *et al.* [1]. We trained a network on the DRIVE dataset and evaluated its performance on five other retinal vasculature segmentation datasets. As in the image segmentation experiment, we compared the performance of a pre-trained network with a baseline trained from scratch.

Moreover, to analyze how much of the performance drop accompanying model transfer can be attributed to domain shift and how much is caused by labeling mismatch, we tested three modifications: selecting the model checkpoint according to its performance on the target dataset, selecting the classification threshold on the target dataset, and a combination of both. This way, we compared the optimal performance that could be achieved with a model trained on a distinct source dataset and the performance that is achieved in practice due to labeling mismatch. Results of this experiment are part of the appendix (Section 3).

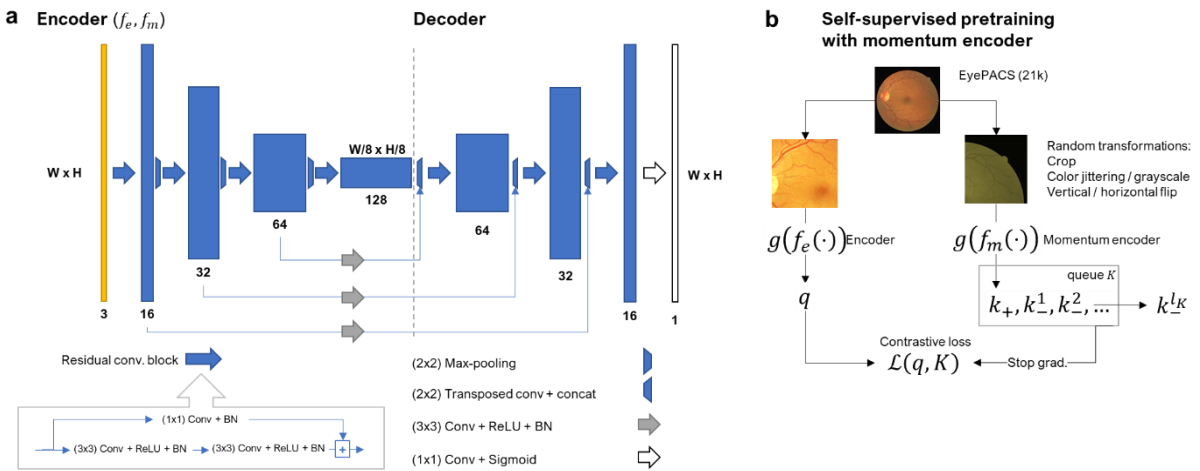

**Fig. 4.1**: Method overview. **a**, Architecture of the used U-Net variant. **b**, Scheme of the momentum contrast (MoCo) self-supervised pre-training approach.

## 4.2 Network architecture

Fig. 4.1a shows a specific variant of U-Net [2] utilized in our experiments. We followed previous research on retinal segmentation which has demonstrated that a simple U-Net can outperform various

other, more intricate networks [1]. Specific architecture details can be found in the appendix (Section 1) or in the provided code.

### 4.3 Self-supervised pre-training

Fig. 4.1b outlines the approach called *momentum contrast* (MoCo) [3] which we used in our experiments for self-supervised encoder pre-training. MoCo is a representation learning method based on the pre-text task of contrastive instance discrimination: randomly transformed variants of training images are encoded by the network to a latent space, where samples originating from the same image should be close to each other and far from samples from other images.

Specifically, images $x_i$ from the training set are transformed by a stochastic transformation $\tilde{t}$ (data augmentation) and encoded to a latent space by an encoder $f_e$ with a non-linear projection head $g$ to obtain a set of keys $k_i \in K$. The same procedure is applied to generate a query sample $q_j$ from an image $x_j$. Cosine distance is used to compute the similarity of $q$ to keys in $K$ and should be small for a positive key $k_+ \equiv k_{i=j}$ and large for other keys $k_{i \neq j}$. Assuming that vectors $q$ and $k$ are $\ell_2$-normalized, this objective can be expressed via the InfoNCE loss function [4, 5] with a temperature hyperparameter $\tau$:

$$\mathcal{L}(q, K) = -\log \frac{\exp(q \cdot k_+ / \tau)}{\sum_{k \in K} \exp(q \cdot k / \tau)}. \tag{1}$$

Having a large set of negative keys is essential for good learning, but recomputing embeddings and gradients of the whole training dataset after each update of the encoder is inefficient. To bypass this issue, MoCo uses small batches but maintains the keys in a queue of length $l_K$. This enables efficient contrastive learning since only a small set of query points is passed through the encoder in each iteration, but a large set of negative keys is available in the queue. A necessary requirement for using such a queue is that the latent representation is not changing too rapidly. To ensure this, MoCo employs a separate *momentum encoder* $f_m$ to embed the keys, whose weights $\theta_m$ slowly follow the query encoder weights $\theta_e$ via the momentum update $\theta_m \leftarrow \alpha \theta_m + (1 - m)\theta_e$, where $m$ is the momentum parameter.

### 4.4 Datasets

Table 4.1 contains a summary of datasets used in our work. For pre-training, we used 21 072 images from the EyePACS dataset [6]. The dataset was collected with different fundus cameras at multiple centers in the US and is representative of large variety of medical conditions and imaging artifacts. Although the dataset comes with labels for diabetic retinopathy classification, we did not use them.

**Table 4.1**: Overview of datasets utilized in our experiments. For DRIVE, HRF, and CHASE-DB, we used the same test splits as proposed by [1]. For IDRiD, we used the official test splits.

| Dataset | Train / Val / Test | W × H | Labels |
| --- | --- | --- | --- |
| EyePACS [6] | 21 072 | 400–5184 × 289–3456 | None |
| DRIVE [7] | 16 / 4 / 20 | 565 × 584 | Blood vessels |
| HRF [8] | 12 / 3 / 30 | 3504 × 2336 | Blood vessels |
| CHASE-DB [9] | 6 / 2 / 20 | 999 × 960 | Blood vessels |
| STARE [10] | 20 | 605 × 700 | Blood vessels |
| LES-AV [11] | 22 | 1144 × 1620 1958 × 2196 | Blood vessels |
| AV-WIDE [12] | 30 | 2816 × 1880 1500 × 900 | Blood vessels |
| DR HAGIS [13] | 39 | 2816 × 1880 4752 × 3168 | Blood vessels |
| IDRiD IDRiD (SE) [14] | 43 / 11 / 27 21 / 5 / 27 | 4288 × 2848 | Retinal lesions, optic disc, fovea |

For blood vessel segmentation, we utilized several datasets: DRIVE [7], HRF [8], CHASE-DB [9], STARE [10], LES-AV [11], AV-WIDE [12], and DR HAGIS [13]. For image segmentation experiments, we followed train/test splits proposed by [1]. For domain transfer experiments, we used the training split of the source dataset and tested on all data from the target dataset. We repeated experiments with different train/validation splits.

Furthermore, we utilized the IDRiD dataset for lesion segmentation evaluation [14]. The dataset comes with segmentations of hard exudates (EX), soft exudates (SE), hemorrhages (HE), microaneurysms (MA), and optic disc (OD). Additionally, the location of the fovea (FO) is provided. The train/test splits are provided by the authors. We reserved 20% of the training data for validation.

### 4.5  Training setup

For pre-training of the encoder, we first resized and cropped the images to a uniform size 512×512 px. Then, we generated two input samples from each image in the mini-batch by applying data augmentation, with one fed to the encoder and the other to the momentum encoder. Samples processed by the momentum encoder were stored in a queue of length 4 096 and used to compute the loss. In total, we pre-trained the network for 600 epochs.

For training the whole U-Net on segmentation tasks, we resized the images from HRF and IDRiD datasets to a width of 1024 px, images from DRIVE and CHASE-DB datasets were used in their original resolution. For IDRiD, we trained networks for each segmentation target separately (one-vs-all). In total, we trained the networks for 1500 epochs, saved checkpoints every 10 epochs, and finally selected the checkpoint with the best Dice score on the validation set. The validation sets were created by reserving 20% of the training data (rounded down) before the training. In experiments with reduced training set, the validation sets were still selected as 20% of the whole training set, not the reduced one. No pre-processing of the images or post-processing of the segmentation masks other than resizing was used.

Additional details can be found in the appendix (Section 1) and in the provided code.

### 4.6  Evaluation protocol

To ensure good reproducibility of our results, we adopted the evaluation protocol described by Galdran *et al.* [1]. In short, we followed the same train/test splits (Table 4.1). We selected the final classifier threshold which maximizes the Dice coefficient on the training set predictions, and evaluated the final performance on all pixels of all testing images (except masked areas outside of the field-of-view) at their original resolution together (contrary to computing mean performance over individual images). Furthermore, we also utilized test-time augmentation by averaging the predicted segmentations for all four possible horizontal and vertical flip combinations.

For blood vessel segmentation tasks, we evaluated the predictions using the Dice coefficient (F1 score). For lesion segmentation, we computed the area under precision-recall curve (AUPRC) to match the evaluation used in the IDRiD challenge.

### 4.7  Statistical analysis

We performed statistical testing of the following hypotheses: in the image segmentation experiment, we tested whether 1) self-supervised pre-training leads to a higher Dice coefficient than the baseline, 2) pre-trained models converge in a lower number of epochs than the baseline, and in the domain transfer experiment, we tested the hypothesis that 3) self-supervised pre-training leads to a higher Dice coefficient on a transfer dataset than the baseline.

We repeated the experiments several times (N=4 for image segmentation, N=12 for domain transfer) and paired the results from matching training/validation splits. To test the hypotheses, we computed one-sided 95% $t$-confidence intervals on the differences and declared the difference significant if zero was outside the CI. This test assumes normality of the distribution of differences, which was validated by the Shapiro-Wilk test.

The analysis was performed using SciPy (v. 1.6.3) [15] and MS Excel (v. 2104).

# Supplementary material

## 1. Network architecture and training setup

**U-Net.** We utilized a U-Net type of network comprising of a fully convolutional encoder-decoder pair with additional skip connections. The encoder has 4 levels, each composed of two blocks of 3x3 convolution, ReLU activation, and batch normalization (BN). Additionally, each level has an additive residual connection with 1x1 convolution and BN. Encoder levels are separated by 2x2 max-pooling layers. The encoder is followed by 3 decoder levels composed of the same blocks as the encoder levels. Before every decoder level, a transposed 2x2 convolution with stride 2 is used, whose output is concatenated to a skip connection from the encoder level with corresponding resolution. Optionally, the skip connections contain a 3x3 convolution, ReLU, and BN. Finally, a 1x1 convolution with sigmoid activation serves as the final binary classifier. The first encoder level uses 16 filters; this number is doubled after each pooling operation and halved by each transposed convolution. One exception is the domain transfer experiment where we use a constrained decoder with 16-8-4 features to reduce overfitting.

**Projection head.** Using a non-linear projection head on top of the encoder improves the representations learned by contrastive self-supervised learning as it avoids enforcing invariance to transformations $\tilde{t}$ [1, 2]. We utilized a shared non-linear projection head $g$ on top of both $f_e$ and $f_m$ consisting of global average pooling and two fully connected layers per 128 units with ReLU activation after the first one.

**Training setup.** Supplementary table 1 summarizes the hyperparameters used during self-supervised pre-training (representation learning experiment). The temperature hyperparameter $\tau$ and the momentum $m$ were set according to previous reports [3, 4]. Supplementary table 2 summarizes hyperparameters used for training of image segmentation networks (image segmentation experiment). Supplementary table 3 summarizes parameters used for cross-training (domain-transfer experiment). Weight initialization followed the scheme proposed by He *et al.* [5] in all experiments.

**Supplementary table 1:** Training hyperparameters for self-supervised encoder pre-training. LR = learning rate.

| | |
|---|---|
| **Dataset** | EyePACS (first 21 072 images) |
| **Data processing / augmentation** | Resize + crop (512x512 px) |
| | Random crop (128x128 px) |
| | Color jitter (probability=80%, brightness 0.4, contrast 0.4, saturation 0.4, hue 0.1) |
| | Grayscale (p=20%) |
| | Horizontal flip (p=50%) |
| | Vertical flip (p=50%) |
| **Batch size** | 64 |
| **Optimization** | Adam, LR cosine schedule from $10^{-2}$ to $10^{-8}$ with restarts after 50 epochs |
| **Weight decay** | $10^{-4}$ |
| **Training length** | 600 epochs |
| **Queue length** ($l_K$) | 4096 |
| **InfoNCE temperature** ($\tau$) | 0.07 |
| **Momentum** ($m$) | 0.999 |

**Supplementary table 2:** Training hyperparameters for segmentation network fine-tuning and training from random initialization. LR = learning rate.

| **Dataset** | DRIVE | HRF | CHASE-DB | IDRiD |
|---|---|---|---|---|
| **Data processing** | | Resize (682x1024 px) | | Resize (680x1024 px) |
| **Data augmentation** | Random rotation [−45°,45°] or scaling [0.95, 1.2] or horizontal translation [−5%, 5%] | | | |
| | Color jitter (brightness 0.25, contrast 0.25, saturation 0.25, hue 0.1) | | | |
| | Horizontal flip (p=50%) | | | |
| | Vertical flip (p=50%) | | | |

| | | |
|---|---|---|
| **Batch size** | 4 | |
| **Optimization** | Adam, LR cosine schedule from $10^{-2}$ to $10^{-8}$ with restarts after 50 epochs | Adam, LR=$10^{-3}$ |
| **Weight decay** | 0 | |
| **Training length** | 1500 epochs | |
| **Convolutional skip connections** | No | Yes |

**Supplementary table 3:** Training and testing hyperparameters for cross-training experiments. LR = learning rate.

| | Training |
|---|---|
| **Dataset** | DRIVE |
| **Data processing** | Resize (512x529 px) |
| **Data augmentation** | Random rotation [−45°,45°] or scaling [0.95, 1.2] or horizontal translation [−5%, 5%]<br>Color jitter (brightness 0.25, contrast 0.25, saturation 0.25, hue 0.1)<br>Horizontal flip (p=50%)<br>Vertical flip (p=50%) |
| **Batch size** | 4 |
| **Optimization** | Adam, LR cosine schedule from $10^{-2}$ to $10^{-8}$ with restarts after 50 epochs |
| **Weight decay** | $10^{-4}$ |
| **Training length** | 1500 epochs |
| **Convolutional skip connections** | No |
| **Decoder layers** | 16-8-4 |

| | Testing | | | | | |
|---|---|---|---|---|---|---|
| **Dataset** | HRF | DR HAGIS | AV-WIDE | LES-AV | CHASE-DB | STARE |
| **Data processing** | Resize 682x1024 px | Resize 682x1024 px | Resize width=1024 px | Resize width=512 px | Resize width=512 px | Resize width=512 px |
| **Augmentation** | Horizontal and vertical flipping | | | | | |

## 2. Training convergence analysis

**Supplementary table 4**: Mean training length improvement of pre-trained network over a randomly initialized U-Net (epoch count). The number of experiment runs considered is listed under N.

| Lesions | N | Difference | 95% CI |
|---|---|---|---|
| Hard exudates | 20 | 232 | 64, $+\infty$ |
| Soft exudates | 16 | 205 | -139, $+\infty$ |
| Microaneurysms | 20 | 292 | 75, $+\infty$ |
| Hemorrhages | 20 | 283 | 133, $+\infty$ |
| **Blood vessels** | | | |
| DRIVE | 24 | 173 | 99, $+\infty$ |
| HRF | 28 | 180 | 50, $+\infty$ |
| CHASE-DB | 16 | 289 | 154, $+\infty$ |

## 3. Domain-transfer error analysis

We observed that although the self-supervised pre-training improves the model transfer, training even on a single image from the target dataset can yield superior performance to transferring a model from another, larger dataset (cf. Fig. 3). This performance gap is partially caused by domain shift, in which case the network is unable to recognize patterns in images from a different dataset, and can be potentially remedied by learning a more robust representation. Additionally, the performance is reduced due to labeling mismatch, in which case the network can recognize patterns, but it was trained to label them differently (e.g., small vessels). This mismatch is expected in fundus photographs, since the presence of retinal blood vessels cannot be exactly represented as a binary mask and there is a considerable inter-observer variability regarding how small vessels and vessel edges are handled [6]. The relevance of labeling-related performance drop for medical segmentation applications has been questioned in the past [7].

We estimate that about 2.5 pp are caused by labeling mismatch and expect that the remaining ~4 pp could be reduced by further improvements to representation learning. To characterize the contribution of labeling mismatch to the performance gap, we used the target dataset for selection of the optimal checkpoint and classification threshold instead of the source dataset. These changes reduce the error due to labeling error but do not alter the training data and thus do not affect the domain shift problem. Supplementary table 5 lists the performance improvements achieved by these modifications.

**Supplementary table 5:** Influence of checkpoint selection and binarization threshold of models trained on DRIVE and tested on HRF. Means of four runs are reported.

| Training set | Checkpoint | Threshold | Dice (%) |
|---|---|---|---|
| DRIVE | DRIVE | DRIVE | 75.06 |
| | DRIVE | HRF | 75.57 |
| | HRF | DRIVE | 77.11 |
| | HRF | HRF | 77.47 |
| HRF | HRF | HRF | 81.38 |

## 4. Abbreviations

| | |
|---|---|
| AUC | Area under receiver operating characteristic curve |
| AUPRC | Area under precision-recall curve |
| CNN | Convolutional neural network |
| MCC | Matthews correlation coefficient |
| EX | Hard exudates |
| SE | Soft exudates |
| OD | Optic disc |
| MA | Microaneurysms |
| HE | Hemorrhages |
| FO | Fovea |
| ReLU | Rectified linear unit |
| MoCo | Momentum contrast |